\documentclass[aps,pre,twocolumn]{revtex4}
\usepackage{graphicx,amsmath,amsfonts,bm,enumerate}
\newcommand{\pa}{\partial}
\newcommand{\tb}[1]{\tilde{\bm{#1}}}
\renewcommand{\=}{\!=\!}
\newcommand{\avg}[1]{\left\langle #1 \right\rangle}
\newcommand{\quen}[1]{\left\langle\!\left\langle #1 \right\rangle\!\right\rangle}

\renewcommand{\t}[1]{\tilde{#1}}
\renewcommand{\O}{\mathcal{O}}
\begin{document}

\title{On the spatial distribution of thermal energy in equilibrium}
\author{Yohai Bar-Sinai and Eran Bouchbinder}
\affiliation{Chemical Physics Department, Weizmann Institute of Science, Rehovot 7610001, Israel}

\begin{abstract}
The equipartition theorem states that in equilibrium thermal energy is equally distributed among uncoupled degrees of freedom which appear quadratically in the system's Hamiltonian. However, for spatially coupled degrees of freedom --- such as interacting particles --- one may speculate that the spatial distribution of thermal energy may differ from the value predicted by equipartition, possibly quite substantially in strongly inhomogeneous/disordered systems. Here we show that for systems undergoing simple Gaussian fluctuations around an equilibrium state, the spatial distribution is universally bounded from above by $\frac{1}{2}k_BT$. We further show that in one-dimensional systems with short-range interactions, the thermal energy is equally partitioned even for coupled degrees of freedom in the thermodynamic limit and that in higher dimensions non-trivial spatial distributions emerge. Some implications are discussed.
\end{abstract}

\maketitle

Equilibrium thermal fluctuations play a key role in physics, chemistry and biology, and the framework that captures their properties --- statistical thermodynamics --- is a central branch of physics. One of the renowned results obtained in this field is the equipartition theorem \cite{huang1963}, which in its simplest form states that the total thermal energy of the system is equally distributed among its uncoupled degrees of freedom (DOF). In addition, each uncoupled DOF appearing quadratically in the Hamiltonian has on average an energy of $\tfrac{1}{2}k_B T$, where $k_B$ is the Boltzmann constant and $T$ is the absolute temperature~\cite{huang1963}.

The equipartition theorem holds only for \emph{uncoupled} DOF, and strictly speaking does not state anything about the energy of coupled DOF. Quadratic Hamiltonians can always be decomposed into a set of uncoupled DOF  (``mutually orthogonal normal modes''), and the theorem applies to them. It is intriguing, though, to ask in all generality whether the average potential energy of coupled DOF can significantly deviate from the value predicted by equipartition. Put simply, we ask: what can be said in general about the spatial distribution of thermal energy? To the best of our knowledge, despite its basic nature and generality, this question has not been posed in the literature, nor answered.

One may speculate that a localized enhancement or inhibition of thermal energy may have some effect on various local processes. For instance, if local thermal fluctuations activate chemical reactions, these might be facilitated or hindered in the presence of enhanced or reduced fluctuations. Other processes which might be affected are structural changes, such as severing of biopolymers~\cite{Elam2013, Ngo2015}. Related effects may also be observed in elastic-network models of protein folding, where local (nearest-neighbor)  fluctuations are assumed to dictate bond rupture~\cite{Srivastava2013}. We note that while much attention has been devoted to the effect of disorder on non-equilibrium transport properties, e.g.~\cite{Casher1971, Rich1975, Livanov1991}, to the best of our knowledge none of the previous works explored the equilibrium spatial distribution of thermal energy, as we intend to do here.
\begin{figure}[h]
 \centering
 \includegraphics[width=0.7\columnwidth]{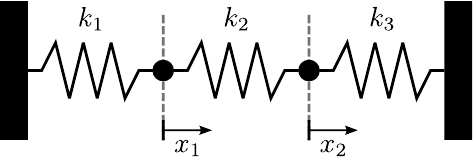}
 \caption{A model system of two masses and three linear springs, connected in series between fixed walls. $x_i$ measures the deviation of the $i$-th particle from its equilibrium position.}
 \label{fig:toy}
\end{figure}

As a prelude, we begin by solving a very simple problem, depicted in Fig. \ref{fig:toy}. Consider a system of two particles interacting via linear springs with each other and with bounding walls. Its potential energy is
\begin{equation}
U = {\textstyle\sum}_{\alpha=1}^3\epsilon_\alpha= \tfrac{1}{2}k_1 x_1^2 + \tfrac{1}{2}k_2 (x_2-x_1)^2 + \tfrac{1}{2}k_3 x_2^2,
\label{eq:springs_energy}
\end{equation}
where $\epsilon_\alpha$ is the energy of the $\alpha$-th spring and the $x_i$'s denote the deviation of the $i$-th particle from its equilibrium position (when possible, we adopt the convention that Latin indices denote DOF, while Greek indices denote interactions). In Eq.~\eqref{eq:springs_energy}, we assume that the rest length of the entire chain is identical to the distance between the bounding walls. As is typical in such systems, the kinetic energy, $\sum_i \tfrac{1}{2}m_i\dot x_i^2$, is a sum of quadratic uncoupled terms. Thus, diagonalizing the kinetic contribution is trivial, and hence in what follows we disregard the kinetic energy of the system.

What is the average energy stored in the $\alpha$-th spring? In particular, can it significantly deviate from the value predicted by equipartition? For such a simple system the answer is readily calculable through the correlations between the $x_i$'s. For example, $\avg{\epsilon_2}$ is given by
\begin{equation*}
 \avg{\epsilon_2}\=\avg{\frac{k_2}{2}(x_2-x_1)^2}\=\frac{k_2}{2}\Big(\avg{x_1^2}+\avg{x_2^2}-2\avg{x_1x_2}\Big)\ ,
\end{equation*}
where $\avg{\cdot}$ denotes thermal averaging. Since the energy is quadratic, the correlation matrix $\bm C$ is given in terms of the Hamiltonian $\bm H$ by~\cite{huang1963}
\begin{equation}
C_{ij} = \avg{x_i x_j} = k_B T ({\bm H}^{-1})_{ij} \ ,
\label{eq:correlation}
\end{equation}
where the the Hamiltonian (or the Hessian) is defined as $ H_{ij}\!\equiv\!\frac{\partial^2 U}{\partial x_i\partial x_j}$. With these formul\ae, an explicit calculation yields~\footnote{Whenever $\bm H$ is non-invertible, i.e.~in the presence of Goldstone modes, the notation $\bm{H}^{-1}$ should be interpreted as the Moore-Penrose pseudoinverse~\cite{Penrose1955, Meyer2000}. See supplementary material for details.}
\begin{equation}
 \avg{\epsilon_\alpha}=\tfrac{1}{2}k_B T\left[1-\frac{k_\alpha^{-1}}{k_1^{-1}+k_2^{-1}+k_3^{-1}}\right] \ .
 \label{eq:epsilon_springs}
\end{equation}

A few insights can be gained from this very simple example. First, a clear cut answer is given to the question presented above: for a general choice of the $k_\alpha$'s, the \textit{thermally averaged} energy of a given spring may differ from the value predicted by equal partition. Second, since the system consists of two DOF (i.e. the Hamiltonian has two normal modes), each contributes $\tfrac{1}{2}k_B T$ to the total energy, and thus $\sum_\alpha\!\avg{\epsilon_\alpha}\=k_B T$, as expected. That is, the \textit{spatial average} of the energy agrees, by construction, with the equipartition theorem, and reads $(N+1)^{-1}\sum_\alpha\!\avg{\epsilon_\alpha}\=\tfrac{1}{3}k_B T$ (with $N\=2$). Third, we note that $\avg{\epsilon_\alpha}$ is bounded between 0 and $\tfrac{1}{2}k_B T$, which means that inhomogeneity in the $k_\alpha$'s might either increase or decrease it relative to the spatially average value of $\tfrac{1}{3}k_B T$, depending on the inhomogeneity. Lastly, we note that for a homogeneous system, i.e. $k_\alpha\=k$, the energy is obviously equally partitioned among the springs and equals $\tfrac{1}{3}k_B T$.

These results might appear somewhat restricted as they involve a system with a small number of DOF, $N\=2$, and involve specific boundary conditions which may play a non-trivial role, especially in a small system. Consequently, we next aim at understanding how the spatial distribution of thermal energy depends on the number of DOF, $N$, and on the boundary conditions.

We first consider a system of $N$ DOF $\bm{x}\=(x_1,\dots,x_N)^T$ and $N+1$ springs. The potential energy is
\begin{equation}
U={\textstyle\sum}_{\alpha=1}^{N+1} \epsilon_\alpha\ , \qquad\epsilon_\alpha=\tfrac{1}{2}k_\alpha\left(x_\alpha-x_{\alpha-1}\right)^2  \ ,
 \label{eq:epsilon}
\end{equation}
where $1\!\le\!\alpha\!\le\!N+1$, $\epsilon_\alpha$ is the potential energy of the $\alpha$-th spring and the $k_\alpha$'s are non-negative constants. Formally, Eq.~\eqref{eq:epsilon} makes reference to $x_0$ or $x_{N+1}$. These are not real DOF, but rather ``boundary conditions'' imposed by the fixed walls and should be taken as $x_0\=x_{N+1}\=0$. The general question that we pose is: what can be stated about the distribution of~$\avg{\epsilon_\alpha}$, for general $k_\alpha$'s?

The conventional procedure for addressing such a question is to diagonalize the Hamiltonian $\bm H$ and work in the basis of its normal modes. Then, $\avg{\epsilon_\alpha}$ can be reconstructed, at least conceptually, by summing over the contributions of the individual modes (e.g. Eq. (9) in \cite{Atilgan2001}). While this generic recipe is very useful in most cases, in this case working with the normal modes obfuscates the structure of the problem, since there is no simple way to describe them for a general distribution of the $k_\alpha$'s.

What is then a useful basis to work with? To answer this question we first note that for \textit{any} linear change of variables $\tb x\!\equiv\!\bm{A\,x}$, where $\bm A$ is an invertible $N\!\times\!N$ matrix, the modified Hamiltonian takes the form $\tb{H}=\bm{A}^{-T}\bm{H}\bm{A}^{-1}$, where $\bm{A}^{-T}$ stands for $(\bm{A}^{-1})^T$ (note that this is not a similarity transformation, as $\bm A$ is not orthogonal). Straightforward matrix manipulations show that even for the non-orthogonal variables $\tb{x}$ the correlation matrix is given by the inverse of the relevant Hamiltonian, i.e. $\tilde{\bm C}\equiv\avg{\tb{x}\tb{x}^T}=k_B T {\tb{H}}^{-1}$.
The main insight gained from this brief discussion is that one should not be constrained to using an orthogonal basis in transforming the Hamiltonian into a desired form. This turns out to be important for solving the problem at hand.

Following this insight, we look for new variables $\tb x\!\equiv\!\bm{A\,x}$ such that $\tb H$ will be, loosely speaking, ``almost'' diagonal. A clue for finding a useful basis is obtained by inspecting Eq. \eqref{eq:epsilon}, which is \textit{already} written in an ``almost'' diagonal form, if we identify the new variables simply as $\t x_i\!\equiv\!x_i-x_{i-1}$. That is, we take the combinations that make up the interactions as the new variables. This defines the transformation matrix $A_{ij}\!\equiv\!\delta_{ij}-\delta_{i,j+1}$.

Under this choice of non-orthogonal variables, almost all of the spring energies in Eq.~\eqref{eq:epsilon} become $\epsilon_\alpha\=\frac{1}{2}k_\alpha\t{x}_\alpha^2$. The last relation, however, is valid only for $1\!\le\!\alpha\!\le\!N$. Clearly, a one-to-one correspondence between the DOF and the interactions/springs is impossible since the number of springs exceeds $N$. Indeed, an explicit calculation shows that in terms of the new variables the energy is not strictly decoupled, but only almost
\begin{equation}
\label{eq:Htilde}
\begin{split}
 U&={\textstyle\sum}_{\alpha=1}^N \tfrac{1}{2}k_\alpha \t x_\alpha^2+\tfrac{1}{2}k_{N+1}\left(\t x_1+\dots+\t x_N\right)^2\\
  \t{H}_{ij}&=k_i \delta_{ij}+k_{N+1}\ ,\quad\hbox{or,}\quad {\tb H}= \bm{K}+k_{N+1}\tb b\tb b^T\ ,
\end{split}
\end{equation}
where $\bm K\!\equiv\!\operatorname{diag}(k_1,\dots,k_N)$ and $\tb b$ is a vector of $1$'s.

Equation \eqref{eq:Htilde} is very useful since the inverse $\tb{H}^{-1}$ is readily calculated using the Sherman-Morrison formula~\cite{Sherman1950}, which can be expressed in the form
\begin{equation}
 \left(\bm H+k\,\bm{v v}^T\right)^{-1}=\bm H^{-1}-\frac{\bm H^{-1}\bm {v v}^T \,\bm H^{-1}}{k^{-1}+\bm{v}^T\bm{H}^{-1}\bm v}
 \label{eq:SM_formula}
\end{equation}
and is valid whenever both $\bm H$ and $\bm H+k\,\bm{vv}^T$ are invertible (here ${\bm v}$ is a vector and $k$ is a scalar). Applying this formula to Eq.~\eqref{eq:Htilde}, we obtain the generalization of Eq.~\eqref{eq:epsilon_springs} to {\em any} $N$
\begin{equation}
 \avg{\epsilon_\alpha}=\tfrac{1}{2}k_B T\left[1-(N+1)^{-1}\frac{k_\alpha^{-1}}{\quen{k^{-1}}}\right] \ ,
 \label{eq:epsilon_springs2}
\end{equation}
where $\quen{k^{-1}}\!\equiv\!(N+1)^{-1}\sum_\alpha k_\alpha^{-1}$ is the quenched average.

Two features of this result will prove important. First, it is seen that $\avg{\epsilon_\alpha}<\tfrac{1}{2}k_BT$ regardless of the choice of $k_\alpha$'s (moreover, the order in which the $k_\alpha$'s are distributed in space makes no difference, $\avg{\epsilon_\alpha}$ depends only on $k_\alpha^{-1}$ and the average $\quen{k^{-1}}$). Second, it is evident that $\avg{\epsilon_\alpha}$ tends towards $\tfrac{1}{2}k_B T$ in the thermodynamic limit $N\!\to\!\infty$, as long as $k_\alpha^{-1}/\quen{k^{-1}}$ does not increase with $N$.

Our next task is to generalize Eq.~\eqref{eq:epsilon_springs2}. It will be shown that these two features are general for a wide class of physical systems, namely systems with local interactions undergoing Gaussian fluctuations around a stress-free equilibrium. As we will show, for these systems $\frac{1}{2}k_B T$ is a \textit{strict upper bound} for $\avg{\epsilon_\alpha}$. This result is entirely general, independent of dimensionality or interaction range. Second, the fact that $\avg{\epsilon_\alpha}\=\frac{1}{2}k_B T$ plus a negative correction of order $N^{-1}$, which depends on the inhomogeneity, is the general rule for one-dimensional systems with short-range interactions. In particular, in such systems the distribution of thermal energy becomes \textit{spatially constant} in the thermodynamic limit. These are two main results of this work.

Consider a system with $N$ DOF and total energy $U\!=\!\sum_{\alpha=1}^n\!\epsilon_\alpha$, where $n$ is the number of interactions. The most general expansion of $\epsilon_\alpha$ in the DOF reads
\begin{align}
 \epsilon_\alpha&=\epsilon^{(\!\alpha)}_0+{\textstyle\sum}_i F^{(\!\alpha)}_i x_i + \tfrac{1}{2}{\textstyle\sum}_{i,j} x_i C^{(\!\alpha)}_{ij}x_j + \mathcal{O}\left(\bm x^3\right) \ .
 \label{eq:epsilon_alpha}
\end{align}
The linear term vanishes under thermal averaging as long as anharmonic contributions to the energy are neglected, and hence is omitted hereafter (note that $\sum_\alpha\! F^{(\!\alpha)}_i\!=\!\pa U/\pa x_i$ vanishes due to global equilibrium). The only assumption we adopt is that ${\textstyle\sum}_{i,j} x_i C^{(\!\alpha)}_{ij}x_j$ can be written as $({\sum}_i B_{\alpha i}\,x_i)^2$, where $\bm B$ is an $n\!\times\!N$ matrix which describes the interactions in the system (for example, $\bm B$ can be easily read off Eq.~\eqref{eq:epsilon}). In this case, the Hamiltonian is given by $\bm{H}\=\bm{B}^T\bm{B}$.

This is a generic form of local interaction energies for a broad class of physical systems: (i) Discrete field theories, or discrete approximations to continuum field theories, where the energy density takes the form $[\mathcal{L}(f)]^2$, with some spatial linear differential operator $\mathcal L$ and field $f$. Relevant examples include --- among many others --- the Euler-Bernoulli theory of elastic beams \cite{LLElasticity}, the F\"oppl von-K\'arm\'an theory of thin sheets~\cite{LLElasticity} and the Helfrich theory of membrane elasticity \cite{Safran1994}. (ii) Systems of discrete particles interacting via a radially-symmetric pairwise potential, where at equilibrium all particle pairs are at a stress-free configuration. Relevant examples include glassy systems near jamming~\cite{Hecke2010} and elastic networks~\cite{Srivastava2013}.

Using Eq.~\eqref{eq:correlation}, $\avg{\epsilon_\alpha}$ can be readily expressed in terms of the interaction matrix~$\bm B$, as
\begin{align}
 \avg{\epsilon_\alpha}&=\tfrac{1}{2}k_B T\,P_{\alpha\alpha},&
 \bm{P}&\equiv\bm B \left(\bm{B}^T\bm{B}\right)^{-1}\bm B^T\ .
 \label{eq:P}
\end{align}
$\bm P$ is an orthogonal projection operator \cite{Meyer2000}, since $\bm P^2\=\bm P$ (this holds even when $\bm{B}^T\bm{B}$ is not invertible~\cite{SM}). A general property of such operators is that all their elements are smaller than unity in absolute value. We thus prove a central result of this work, i.e.~that
\begin{align}
 \avg{\epsilon_\alpha}=\tfrac{1}{2}k_B T\,P_{\alpha\alpha}\le \tfrac{1}{2}k_BT\ .
 \label{eq:upper_bound}
\end{align}

The rank of $\bm P$, which equals that of $\bm H$ and $\bm B$, carries important information. For example, since $\bm P$ is a projection operator that works in a space of dimension $n$, in case that $\operatorname{rank}\bm P\=n$ we can immediately conclude that $\bm P$ is the identity and thus $\avg{\epsilon_\alpha}=\frac{1}{2}k_BT$ identically. This happens whenever the rows of the interaction matrix $\bm B$ are linearly independent (and in particular $n\!\le\!N$). The case $n\!=\!N$, for nearest-neighbor interactions, corresponds to isostatic systems, i.e. systems where the number of constraints (interactions) equals the number of DOF~\cite{Hecke2010}.

Equation \eqref{eq:upper_bound} puts strict bounds on the possible values of $\avg{\epsilon_\alpha}$, but much more can be said about the behavior within these bounds. Specifically, we can derive the analog of Eq.~\eqref{eq:epsilon_springs2} in the general case of one-dimensional systems with short range interactions. The detailed derivation can be found in~\cite{SM}, and it follows verbatim the structure of the derivation of Eq.~\eqref{eq:epsilon_springs2}, as outlined here: For one-dimensional systems, where each DOF interacts with its $m$ nearest neighbors the number of interactions generally exceeds the number of DOF by $m$. A non-orthogonal transformation is used to bring the Hamiltonian to the form $\tb{H}=\bm K+\sum_\alpha k_\alpha\tb{b}_\alpha\tb{b}_\alpha^T$, where $\bm K$ is diagonal and the second term is a sum over the $m$ ``excess'' interactions. Then, the Sherman-Morrison formula is iteratively applied $m$ times to calculate the inverse. Because of the short-range nature of the interactions, the magnitude of the non-diagonal correction to $\tb{H}^{-1}$ is of order $m/N$, and vanishes in the thermodynamic limit. Thus, $\avg{\epsilon_\alpha}=\frac{1}{2}k_BT+\mathcal{O}(N^{-1})$.

This is another central result of this work: in one-dimensional systems with short range interactions, the spatial distribution of thermal energy becomes essentially flat in the thermodynamic limit. The crux of the argument lies in the fact that the number of interactions does not greatly exceeds the number of DOF, and that the ratio between them approaches unity in the the thermodynamic limit.

Returning to the problem of the spring-mass chain, and having solved the problem of the $N$-dependence for fixed boundary conditions, we now turn to explore the effect of boundary conditions for a fixed $N$. For instance, semi-fixed boundary conditions can be obtained by removing the constraint of one of the walls, by setting, say, $k_1\=0$. In doing so we obtain $n\=N$ independent interactions and thus $\avg{\epsilon_\alpha}\=\tfrac{1}{2}k_B T$ is \textit{identically constant} for {\em any} $N$. Fully free boundary conditions are obtained by setting both $k_1$ and $k_{N+1}$ to zero, and give rise to a single Goldstone mode (uniform translation). In this case we have $n\=N-1$ independent interactions and again $\avg{\epsilon_\alpha}$ is identically constant (clearly, both these results can also be obtained by properly taking limits of Eq.~\eqref{eq:epsilon_springs2}). This also shows that in one-dimensional systems the effect of boundary conditions in non-local, i.e. every spring in the system is affected by the bounding walls.

The general approach discussed above can be applied to different types of interactions. For example, bending fluctuations are described by local interactions of the form $\epsilon_\alpha\=\frac{1}{2}\kappa_\alpha\left(x_{\alpha-1}\!-\!2x_\alpha\!+\!x_{\alpha+1}\right)^2$, where the $\kappa_\alpha$'s are the local bending rigidities. Identical arguments show that in a chain with free boundary conditions, the spatial distribution of bending fluctuational energy is exactly constant, regardless of the choice of the $\kappa_\alpha$'s.

This result offers the first application of the theoretical development described in this paper, as it seems to refute a recently conjectured mechanism for severing of actin filaments, one of the most important and ubiquitous biopolymers in eukaryotic cells \cite{Bravo-Cordero2013}. In~\cite{McCullough2008, DeLaCruz2009, McCullough2011}, it has been hypothesized that thermal energy may be concentrated at the boundaries between relatively softer and stiffer regions of the biopolymer (softening is induced by a different molecule, cofilin, which partially binds actin~\cite{Prochniewicz2005, McCullough2008}), and that the excess thermal energy is responsible for the experimentally observed preferential severing near these boundaries. Our result shows, at least within the framework of a discrete description of quadratic bending fluctuations, that no such energy concentration takes place.

What happens in higher dimensions? As the crux of the argument lies in the relative number of DOF and interactions, dimensionality appears to be crucial. In fact, in dimensions higher than one the argument seems to fail qualitatively as generally there are significantly more interactions than DOF. In this case, we also expect local topological variations, bond strength disorder, defects, holes, free boundaries and the like to play a role.

\begin{figure}[t]
 \centering
 \includegraphics[width=\columnwidth]{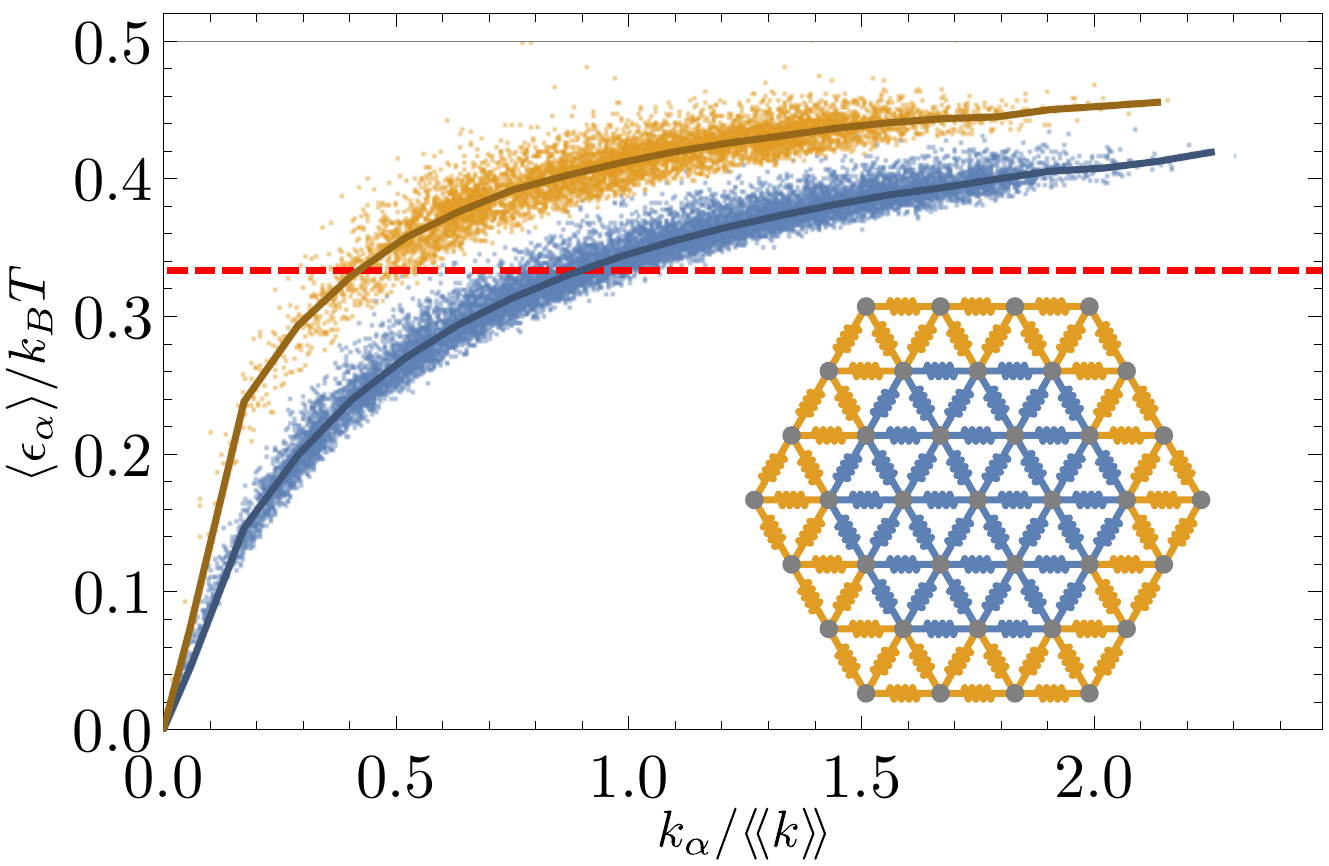}
 \caption{(Color online) The average thermal energy $\avg{\epsilon_\alpha}$ (in units of $k_BT$) as a function of $k_\alpha/\quen k$, for a hexagonal portion of a two-dimensional triangular lattice. The data are partitioned into bulk (blue) and boundary springs (yellow), cf. inset. Spring constants are distributed normally with mean $1$ and variance $0.3$. Each side of the hexagon consists of $20$ springs, which means $N\!=\!2522$ (a smaller system, with $3$ springs and $N\!=\!74$, is shown in the inset for illustration). No quantitative change was observed with increasing $N$. The points show data from $30$ realizations and the solid lines are guides to the eye. The dashed line shows the asymptotic value $\tfrac{1}{3}$, corresponding to $\avg{\epsilon_\alpha}$ in the bulk of a homogeneous triangular lattice in the thermodynamic limit.}
 \label{fig:2D}
\end{figure}

To see this, consider a hexagonal portion of a two-dimensional triangular lattice of $\tfrac{1}{2}N$ particles (which amounts to $N$ DOF), interacting via linear springs, as shown in the inset Fig.~\ref{fig:2D}. Clearly, in the limit of large systems the number of springs $n$ approaches $\tfrac{3}{2}N$. When all of the springs are identical, i.e. with no bond strength disorder, we expect the average energy of a spring far from the free boundary to approach $\avg{\epsilon_\alpha}\!\approx\!\tfrac{1}{2}k_B T\!\times\! \frac{N}{n}\!\to\! \tfrac{1}{3}k_B T$ (this was verified by an explicit calculation). There is no reason, however, to expect the thermal energy to be spatially uniform in the presence of inhomogeneities, either in the lattice topology or in the bond strength.

To test this, we considered the lattice in the inset of Fig.~\ref{fig:2D} with bond strength disorder, where the $k_\alpha$'s are normally distributed~\footnote{To keep the springs constants positive, we used $k_\alpha\!=\!\operatorname{max}(\kappa _i,\kappa _m)$, where $\kappa _i$ are normally distributed, and $\kappa _m\!>\!0$ is a small cutoff that has little influence on the results.}. $\avg{\epsilon_\alpha}$ is plotted vs. $k_\alpha$ in the main panel, where the average energy of bulk springs in a homogeneous system, $\tfrac{1}{3}k_B T$, is shown as well. Bulk and boundary springs are distinguished. Several key observations can be made: (i) Unlike in one-dimensional systems, thermal energy spans the whole interval between $0$ and $\tfrac{1}{2}k_B T$, both below and above the homogeneous system bulk value, $\tfrac{1}{3}k_B T$. (ii) $\avg{\epsilon_\alpha}$ appears to vary systematically with the local spring strength $k_\alpha$. (iii) Boundary springs have higher energy than bulk springs. This is a purely topological effect in which boundary springs have less neighbors than bulk springs, an effect that persists near free boundaries in fully ordered systems. In general, disordered systems (e.g. glassy ones \cite{Hecke2010}) feature also bulk topological disorder.

In summary, in this work we posed a basic question in statistical physics: What is the spatial distribution of thermal energy in equilibrium?
We showed that under the stated conditions it is strictly bounded between $0$ and $\frac{1}{2}k_B T$, and that for one-dimensional systems with short-range interactions, the spatial distribution of thermal energy becomes essentially flat in thermodynamic limit, even for highly disordered systems. The crux of the derivation lies in the fact that in one-dimensional systems the number of interactions is the same as the number of DOF, up to an additive constant which is negligible in the thermodynamic limit. In higher dimensions this does not hold, as was explicitly demonstrated in a specific example. Systematically unraveling the relations between the spatial energy distribution and dimensionality, the system's geometry and the form of disorder is a theoretical challenge for future work.

The most outstanding question that emerges from this work is what the influence of the spatial distribution of thermal energy on various physical processes and quantities might be. If local energy fluctuations can affect local process, as was suggested --- for example --- in the context of bond rupture in elastic-network models of protein folding~\cite{Srivastava2013}, then one can imagine the possibility of tailoring high dimensional systems in order to enhance or reduce thermal fluctuations in defined locations, to control local processes of interest. Another important future direction would be to explore the roles played by stresses (both internal and external).

\newpage

\textit{Acknowledgments} We are grateful to E. M. De La Cruz for indirectly introducing us to the questions discussed in this paper and to E. Lerner for pointing out the possible effects of internal stresses. We thank E. Brener, N. Gov, D. Mukamel, S. Safran and M. Urbakh for fruitful discussions. E. B. acknowledges support from the Israel Science Foundation (Grant No. 712/12), the Harold Perlman Family Foundation and the William Z. and Eda Bess Novick Young Scientist Fund.


\onecolumngrid
\newpage
\begin{center}
\textbf{\large Supplemental Materials for:\\``On the spatial distribution of thermal energy in equilibrium''}
\end{center}
 \twocolumngrid
 
\setcounter{equation}{0}
\setcounter{figure}{0}
\setcounter{table}{0}
\setcounter{page}{1}
\makeatletter
\renewcommand{\theequation}{S\arabic{equation}}
\renewcommand{\thefigure}{S\arabic{figure}}
\renewcommand*{\thepage}{S\arabic{page}}
\renewcommand{\bibnumfmt}[1]{[S#1]}
\renewcommand{\citenumfont}[1]{S#1}

%

This short document is meant to provide additional technical details in relation to results appearing in the manuscript.

\section{General formalism}

The general description of the problem addressed in the manuscript is as follows. Consider a system of $N$ degrees of freedom $\bm{x}\=(x_1,\dots,x_N)^T$, and $n$ local quadratic interactions between the DOF,
\begin{align}
 \epsilon_\alpha&=\frac{k_\alpha}{2}\left(\sum_{i=1}^N D_{\alpha i}x_i\right)^2
 =\frac{k_\alpha}{2}\left(\bm d_\alpha\cdot \bm  x\right)^2\ , & 1&\le i \le n
 \label{eq:epsilon_SM}
\end{align}
where $\bm D$ is a dimensionless $n\!\times\!N$ matrix whose $\alpha$-th row is $\bm d_\alpha$, $k_\alpha$ are non-negative numbers with dimensions of energy and $\bm{x}$ is dimensionless. The non-zero elements of $\bm D$ are assumed to be of order unity. Note that this notation and units are slightly different compared to the manuscript as our aim here is to remain as general as possible. The results reported in the manuscript are recovered when we identify $\bm b_\alpha\=\sqrt{k_\alpha}\,\bm d_\alpha$ and ${\bm B}\=\sqrt{{\bm K}^{(n)}}{\bm D}$, where $K^{(n)}_{\alpha\beta}\!\equiv\! k_\alpha\,\delta_{\alpha\beta}$ is an $n\times n$ diagonal matrix. Following the discussion below Eq.~(8) in the main text, the linear term of Eq.~\eqref{eq:epsilon_SM} is omitted.

The total energy is obtained by summing over $\epsilon_\alpha$, and the Hamiltonian can be expressed in terms of $\bm D$ as
\begin{equation}
\bm H\=\bm{D}^T\bm{K}^{(n)}\bm{D}=\sum_{\alpha=1}^n k_\alpha\,\bm{d}_\alpha\bm{d}_\alpha^T\ .
\end{equation}
As described in the main text, $\avg{\epsilon_\alpha}$ is determined by the diagonal elements of the projection operator
\begin{equation}
 \bm P\equiv \bm K^{(n)}\bm D \left(\bm{D}^T\bm{K}^{(n)}\bm{D}\right)^{\dagger}\bm D^T\ ,
\end{equation}
where $\dagger$ stands for the Moore-Penrose pseudo-inverse, see Sect.~\ref{sec:Penrose} for definitions and details. For an invertible matrix, $\dagger$ is simply the inverse.

In order to proceed, there are two distinct cases to examine:
\begin{enumerate}[(i)]
 \item $\{\bm d_\alpha\}$ are linearly independent (and hence $n\!\le\!N$).
 \item $\{\bm d_\alpha\}$ are linearly dependent.
\end{enumerate}

Case (i) is relatively simple. In this case the rank of $\bm B$ is $n$, and so is the rank of $\bm P$. Since $\bm P$ is a projection operator in an $n$-dimensional space, we conclude that $\bm P$ must be the identity operator, and thus $\avg{\epsilon_\alpha}=\frac{1}{2}k_BT$.

It might be useful to explicitly construct the transformation matrix $\bm A$ that shows how this happens. By the rank-nullity theorem we know that there are $N-n$ vectors $\bm{g}_{n+1},\dots,\bm{g}_{N}$ (``Goldstone modes'') which satisfy $\bm{Dg}_\alpha\=\bm{0}$ and do not have an energetic cost. The $n$ rows of $\bm D$ together with the $N-n$ Goldstone modes form a basis, i.e. the new variables take the form ${\tb x}\=\bm A \bm x$ with
\newcommand{\stunt}{\rule{7mm}{0.4pt}}
\begin{equation}
\begin{split}
 \bm{A}&=\begin{pmatrix}
    \stunt & \bm d_{1} & \stunt\\
   & \cdot &\\
   \stunt & \bm d_n & \stunt\\
    \stunt & \bm g_{n+1} & \stunt\\
   & \cdot &\\
   \stunt & \bm g_N & \stunt\\
 \end{pmatrix} \ .
\end{split}
\end{equation}
In these terms, the energy takes the form
\begin{equation}
 U=\sum_{\alpha=1}^n \frac{1}{2}k_\alpha\t{x}_\alpha^2 = \frac{1}{2}\tb x^T \tb H \tb x\ , \qquad
 \tb{H} = \begin{pmatrix}\bm{K}^{(n)} & \\ & \bm{0} \end{pmatrix}\ .
\end{equation}
With this at hand it is immediate to see that
\begin{align}
 \avg{\epsilon_\alpha}=\frac{k_\alpha}{2}\avg{\t{x}_\alpha^2}=\frac{k_\alpha}{2}k_BT\left(\t{H}^\dagger\right)_{\alpha\alpha}=\frac{1}{2}k_BT \ ,
\end{align}
since
\begin{equation}
\tb{H}^\dagger = \begin{pmatrix}{\bm{K}^{(n)}}^{-1} & \\ & \bm{0} \end{pmatrix}\ ,
\end{equation}
where ${K^{(n)}}^{-1}_{\alpha\beta}\!\equiv\! k^{-1}_\alpha\,\delta_{\alpha\beta}$.

We thus conclude case (i) with the following outcome: whenever the interactions (either $\bm b_\alpha$ or $\bm d_\alpha$) are linearly independent, all of the local energies are equal, $\avg{\epsilon_\alpha}=\frac{1}{2}k_BT$, irrespective of any other property of the interactions. The total energy of the system is then $\frac{1}{2}n k_BT$, which amounts to assigning $\frac{1}{2}k_BT$ to every non-Goldstone mode.

Case (ii) is more involved in the sense that it is difficult to make further progress without substantially constraining the structure of the interactions. As will be shown, in order for the result to remain valid (at least in the thermodynamic limit) the number of interactions must not greatly exceed the number of DOF. This is always the case in one-dimensional systems with short range interactions, as we now show.

In the following we consider a one-dimensional systems with local interactions of the form
\begin{align}
 \epsilon_\alpha&=\frac{k_\alpha}{2}\left(\sum_{j=0}^{m} c_j x_{\alpha-j}\right)^2\ ,
 &  1\le \alpha \le N+m
\end{align}
where $m\!\in\!\mathbb{N}$ is the interaction length. Since we are in one-dimension, in this case we have $n\=N+m$. We remind the reader that this is a shorthand writing where the summand might formally make reference to $x_i$ with $i\!<\!0$ or $i\!>\!N$. These are to be understood as externally specified ``boundary conditions'', and should be taken as 0. For simplicity, we assume here that $\bm H$ is invertible, but this assumption is not essential (see Sect.~\ref{sec:Penrose}).

For these interactions, it is guaranteed that $\bm d_1,\dots,\bm d_N$ are linearly independent, and we can thus define
\begin{align}
 \t{x}_\alpha&=\bm{d}_\alpha\cdot\bm{x} & &1\le \alpha \le N \ ,
\end{align}
or equivalently $\tb{x}=\bm{D}_0\bm x$, where $\bm{D}_0$ is the invertible matrix composed of the first $N$ rows of $\bm D$. In terms of the new variables, the Hamiltonian takes the form
\begin{equation}
\begin{split}
 \tb{H}&=\bm K^{(N)}+\sum_{\alpha=N+1}^n k_\alpha\, \bm{D}_0^{-T}\bm{d}_\alpha \bm{d}_\alpha^T\bm{D}_0^{-1}\\
 &=\bm K^{(N)}+\sum_{\alpha=N+1}^n k_\alpha\, \tb{d}_\alpha \tb{d}_\alpha^T \ ,
\end{split}
\label{eq:H_tilde_SM}
\end{equation}
with the notation $\tb{d}_\alpha\!\equiv\!\bm{D}_0^{-T}\bm{d}_\alpha$. Note that here $K^{(N)}_{\alpha\beta}\!\equiv\!k_\alpha\delta_{\alpha\beta}$ is a $N\!\times\!N$ matrix.

It is seen that $\tb{H}$ is a diagonal matrix, supplemented with $m$ non-diagonal corrections (recall that $n\=N+m$). Since in the absence of these corrections the result $\avg{\epsilon_\alpha}\=\frac{1}{2}k_BT$ follows immediately (for $1\!\le\!\alpha\!\le\!N$), it is left only to verify that they do not significantly affect $\bm{H}^{-1}$. To this end, we iteratively make use of the Sherman-Morrison formula \cite{Sherman1950_SM} which can be put in the form
\begin{equation}
 \left(\bm H+k\,\bm{vv}^T\right)^{-1}=\bm H^{-1}-\frac{\bm H^{-1}\bm {v v}^T \,\bm H^{-1}}{k^{-1}+\bm{v}^T\bm{H}^{-1}\bm v}\ ,
 \label{eq:SM_formula_SM}
\end{equation}
and is valid whenever both $\bm H$ and $\bm H+k\,\bm{vv}^T$ are invertible (here ${\bm v}$ is a vector and $k$ is a scalar). To calculate the inverse of $\tb H$, as given in Eq.~\eqref{eq:H_tilde_SM}, we iteratively apply the formula $m$ times, as follows. We define $\tb H_0=\bm K^{(N)}$, and
\begin{equation}
 \tb H_{\alpha}=\tb{H}_{\alpha-1}+k_{N+\alpha}\, \tb{d}_{N+\alpha}\tb{d}_{N+\alpha}^T\qquad 1\le \alpha \le m\ .
\end{equation}
Each iteration, $\tb H_{\alpha+1}^{-1}$ is calculated in terms of $\tb H_{\alpha}$ and $\tb{d}_{\alpha+1}$ using Eq.~\eqref{eq:SM_formula_SM}.

We now turn to estimate the corrections to the inverse, i.e.~the last term in Eq.~\eqref{eq:SM_formula_SM} in each iteration.  Since $\bm D_0$ is a sparse matrix, having non-zero elements only on $m$ diagonals, $\bm{D}_0^{-1}$ is generically a dense matrix. In fact, in the relatively simple case discussed here, a closed form recursive expression can be obtained, reading $\left(\bm D_0^{-1}\right)_{ij}=f(i-j)$, with
\begin{equation}
\begin{split}
  f(z)&=\begin{cases}
        0 & z<0 \\
        c_0 & z=0\\
        -c_0^{-1}\sum_{i=1}^m f(z-k)c_k & z>0 \ ,
       \end{cases}
\end{split}
\end{equation}
as can be readily verified by explicit calculation. Thus, the number of non-zero elements of $\tb{d}_\alpha$ scales with $N$ (as opposed to $\bm d_\alpha$, for which this number is exactly $m$). Explicitly, the $(\alpha,\beta)$ element of the non-diagonal correction in the first iteration takes the form
\begin{equation*}
 \frac{k_\alpha^{-1}k_\beta^{-1}\t{d}_{\alpha}\t{d}_{\beta}}{k_{N+1}^{-1}+\sum_{\gamma=1}^Nk_\gamma^{-1}\t{d}_{\gamma}\t{d}_{\gamma}}\ .
\end{equation*}
where $\t{d}_{\alpha}$ is shorthand for the $\alpha$-th component of $\tb{d}_{N+1}$, which is of order unity. Note that no summation over $\alpha,\beta$ is implied. The above expression shows that the denominator of this correction contains $N+1$ terms of order $\quen{k^{-1}}$ and hence it scales as $N\quen{k^{-1}}$. The numerator, obviously, has no $N$ dependence.
Consequently, we have
\begin{equation}
 \tb{H}_1^{-1}\=\bm K^{-1}+\frac{\bm{\Delta}}{N}+\O(N^{-2})
\end{equation}
where $\bm\Delta$ is a matrix whose elements are of order $k_\alpha^{-1}k_\beta^{-1}/\quen{k^{-1}}$. This argument can be similarly applied in all $m$ iterations, as long as $m\!\ll\!N$, which is always valid in the thermodynamic limit. We can thus conclude that the correlation matrix is
\begin{equation}
 \bm {C}=k_BT\tb{H}^{-1}\=k_BT\,\bm K^{-1}+\O(N^{-1})\ ,
\end{equation}
and that the average value of the local energies is
\begin{equation}
 \avg{\epsilon_\alpha}=\frac{k_\alpha}{2} k_BT\left(\tb{H}^{-1}\right)_{\alpha\alpha}=\frac{1}{2}k_BT+\O(N^{-1})\ .
\end{equation}

\section{Sum over non-Goldstone modes}
\label{sec:Penrose}

In the manuscript, and in the previous section, it was stated that calculating correlations by summing over non-Goldstone modes is equivalent to using the Moore-Penrose inverse \cite{Penrose1955_SM, Ben-Israel_SM, Meyer2000_SM} of $\bm H$. This follows trivially from the definition of the latter.

The Moore-Penrose pseudo-inverse is a generalization of the usual inverse of a matrix. For a symmetric square $N\times N$ matrix $\bm A$, the Moore-Penrose inverse --- denoted by $\bm A^\dagger$ --- can be explicitly calculated in terms of the eigenvectors of $\bm A$~\cite{Ben-Israel_SM, Meyer2000_SM} according to
\begin{equation}
 A^\dagger_{ij}=\sum_{\lambda_\alpha \ne 0}\frac{Q_{\alpha i}Q_{\alpha j}}{\lambda_\alpha}\ ,
 \label{eq:pseudo_SM}
\end{equation}
where $\bm Q$ diagonalizes $\bm A$, in the sense that $\bm{QHQ}^T$ is diagonal and the $i$-th row of $\bm Q$ is an eigenvector with an associated eigenvalue $\lambda_i$.

This formula might appear more transparent in bra-ket notation, in which it reads
\begin{equation}
 \bm{A}^\dagger =\sum_{\lambda_\alpha \ne 0}\frac{\left|q_\alpha\right\rangle \left\langle q_\alpha\right|}{\lambda_\alpha}\ ,
\end{equation}
where the $\left|q_\alpha\right\rangle$'s form an orthonormal basis (i.e. the rows of $\bm Q$). It is easy to verify by explicit calculation that this formula satisfies the four equations defining the pseudo-inverse~\cite[page 40]{Ben-Israel_SM}. In this writing it is also clear that when $\bm A$ is invertible, and hence all of the $\lambda_i$'s are non-zero, we have $\bm A^\dagger=\bm A^{-1}$.

Consider now a system of $N$ degrees of freedom, $\bm{x}\=(x_1,\dots,x_N)^T$, with an Hamiltonian $\bm H$. We work in the basis of normal modes, and diagonalize $\bm H$ with an orthonormal matrix $\bm Q$, in the sense defined above. The eigenmodes $\bm q_1,\dots,\bm q_N$ are the rows of $\bm Q$. The equipartition theorem \cite{Huang1963_SM} says that
\begin{equation}
 \avg{\bm q_i\cdot \bm q_j}=\frac{k_BT}{\lambda_i}\delta_{ij}\ ,
\end{equation}
whenever $\lambda_i\ne0$. From this, the correlation matrix $\bm C$ of the original DOF can be calculated by summing over the eigenmodes
\begin{equation}
  \avg{x_i x_j}=k_BT \sum_{k,l}Q_{k i} \frac{\delta_{kl}}{\lambda_{k}} Q_{l j} \ .
\end{equation}
In the case where $\bm H$ is invertible, we immediately identify $\delta_{\alpha\beta}/\lambda_{\alpha}$ with (the elements of) $\bm{Q}\bm{H}^{-1}\bm{Q}^T$, and thus Eq.~(2) in the manuscript is obtained. In case some of the $\lambda_i$'s vanish, the sum is preformed only on the non-vanishing ones and the result identifies with Eq.~\eqref{eq:pseudo_SM}.

\end{document}